\documentstyle[12pt]{article}
\title{Chiral and axial anomalies in the framework of generalized
Hamiltonian BFV-quantization}
\author{I.Yu.Karataeva and S.L.Lyakhovich\\Tomsk State
University, Tomsk 634010, Russia}
\date{}
\begin{document}
\maketitle
\begin{abstract}
     The regularization scheme is proposed for the constrained
Hamiltonian formulation of the gauge fields coupled to the chiral or
axial fermions. The Schwinger terms in the regularized operator
first-class constraint algebra  are shown to be consistent with the
covariant divergence anomaly of the corresponding current. Regularized
quantum master  equations  are studied, and the Schwinger terms are
found out to break down both nilpotency of the BRST-charge and its
conservation law. Wess-Zumino consistency conditions are studied for
the BRST anomaly and they are shown to contradict to the covariant
Schwinger terms in the BRST algebra.
\end{abstract}
\section{Introduction}
The method of the generalized canonical BFV-quantization
\cite{batvil,batfra,batfrad} is the most consistent general approach
to quantization of the first-class constrained theories. In principle,
it can also provide a tool for studying gauge anomalies, which is, in fact,
widely applied in two-dimensional theories. In particular the method
allows to find the critical parameters in such models as bosonic and
fermionic strings \cite{hwa,katoga,oht}, W-gravity \cite{thi},
two-dimensional $\sigma$-models \cite{fubmahronven} and others.
The attractive feature of the approach is also in its universal
character that makes it possible to describe a full set of the
anomalies in a theory with a pair of the BFV generating equations;
moreover these principle equations are of an identical structure for
all equal rank gauge theories (see in \cite{batfrad}). However, until
recently there were no examples of describing anomalies in the framework
of the operator Hamiltonian BFV-method in the case of $d>2$. This is seemed
to be induced partly by a known belief that "the canonical approach to
quantization is theoretically useful but not practical" in particular
due to the explicit relativistic noncovariance. It should be
also mentioned that consistent regularization problem of equal-time
commutators does really exist in Hamiltonian approach in $d>2$ case, while
it is not actual for $d=2$ because the covariant OPE-technique can be used
in the case.

We describe in the present paper the chiral nonabelian and axial abelian
massless fermion anomalies \cite{jac,zum} in the framework of Hamiltonian
BFV-quantization approach in $d=4$. To do this uniformly we reformulate the
global symmetry as a gauge one for both of the cases by introducing
auxiliary fields which could
be treated as analog of the Stuckelberg variables in a massive theory. As
a result the both symmetries are described by the BRST operator equations of
the similar structure. Then we are proposing an explicitly covariant
regularization scheme for the operator constrained Hamiltonian formalism
and finding the Schwinger terms in the involution relations.
These Schwinger terms induce the breakdown both of the BRST charge
nilpotency and its conservation law, thereby the anomaly becomes of the BRST
form. The mode of operation does not have a specific character for the both
chiral and axial anomalies in its key points and the proposed procedure may
serve as a possible tool in a study of other anomalies in $d>2$.

Let us mention that the Schwinger terms turn out to be consistent with the
covariant anomaly divergence of the corresponding current but they break
down Wess-Zumino consistency condition for the operator BRST-algebra. It is
worth noting that the contradiction between covariance and Wess-Zumino
condition is the usual phenomenon for an anomalous theory itself (see
\cite{fuji}) and for the BRST-anomaly in particular. So,
in an anomalous theory one usually has a difficult choice
between the covariance and the Jacoby identity. We prefer the covariance
for the following reason. We believe that the physically sensible theory
should provide gauge anomalies to be mutually cancelled and thus it must
have the nilpotent and conserved BRST charge (as far as an anomalous theory
does not meet today any satisfactory physical interpretation). This mutual
cancellation should be actual for an arbitrary coordinate system; thus
we seemingly have to describe covariantly each of the anomalies, while the
consistency condition could be provided for the total BRST charge as
far as it becomes nilpotent in each coordinate system. Thus we may treat
the considered theory as a part of a more wide one, where the proposed
description of the anomaly could be applied for the anomaly cancelation
covariant control.

Let us begin with a suitable (for our goal) outline of the symmetries in
theory of the massless fermions chiraly coupled to nonabelian massless
vector field. The theory is described by the action
\begin{equation}
S=\int^{}_{}d^4x\left[i\bar{\psi}\gamma^\mu\nabla_\mu \psi-\frac{1}{4}
{F^a}^{\mu\nu}{F^a}_{\mu\nu}\right].
\end{equation}
(hereafter we use the following notations:
$\alpha,\beta,\mu,\nu,\rho,\sigma=0,1,2,3$;
${\scriptstyle{i,j,k,l}}=1,2,3$; $\eta_{\mu\nu}=diag(-+++)$;
$\gamma^\mu\gamma^\nu+\gamma^\nu\gamma^\mu=2\eta^{\mu\nu}{\bf 1}$,
$\gamma^5\equiv i\gamma^0\gamma^1\gamma^2\gamma^3$; $\gamma^5\gamma^5=1$;
$\gamma^\mu\gamma^5+\gamma^5\gamma^\mu=0$; $\varepsilon^{0123}=1$, $t^a$
--anti-Hermitian normalized basis in the Lee algebra of symmetry
group, \(f^{abc}\) -- corresponding totally anti-symmetrical structural
constants:
${\left[ t^a,t^b\right]}_-=f^{abc}t^c$, $Sp\left[ t^a,t^b\right]=
-\frac{1}{2}\delta^{ab}$.
${A^a}_\nu$ and ${F^b}_{\mu\nu}$
--- gauge vector field and strength tensor:
${F^a}_{\mu\nu}=\partial_\mu{A^a}_\nu-\partial_\mu{A^a}_\nu-
ief^{abc}{A^b}_\mu{A^c}_\nu$;
${{\tilde{F}}^a}_{\mu\nu}=\varepsilon_{\mu\nu\alpha\beta}F^{a\alpha\beta}$,
$A_\nu={A^a}_\nu t^a$, $F_{\mu\nu}={F^a}_{\mu\nu}t^a$).

If the fermion connection in the covariant derivative involves
nonabelian vector field only
\begin{eqnarray}
& &\nabla_\mu\psi=\partial_\mu\psi-i\Gamma_\mu\psi, \quad
\nabla_\mu\bar{\psi}=\partial_\mu \bar{\psi}+i\bar{\psi}{\bar{\Gamma}}_\mu\\
& &\Gamma_\mu=eA_\mu P_l, \quad {\bar{\Gamma}}_\mu=eA_\mu P_r, \quad
P_{\stackrel{{\scriptstyle{l}}}{{\scriptstyle{r}}}} =
\frac{1}{2}\left(1\mp\gamma^5\right),
\end{eqnarray}
so the theory (1.1) with the connection (1.3) has the gauge symmetry,
which is connected with the local transformations of the left fermions,
and the global invariance, which is connected to the transformations of
the right fermions.
\begin{eqnarray}
& &\delta_\varepsilon\psi=ie\varepsilon P_l\psi, \quad
\delta_\varepsilon\bar{\psi} =-ie\bar{\psi} P_r, \quad
\varepsilon=\varepsilon^a(x)t^a.\nonumber\\
& &\delta_\varepsilon{A^a}_\mu=\partial_\mu\varepsilon^a+
ief^{abc}\varepsilon^b{A^c}_\mu
\end{eqnarray}
The following Noether identities corresponding to the symmetries (1.4)
hold off shell:
\begin{displaymath}
ie\left[ \frac{\delta S}{\delta\psi(x)} t^aP_l\psi(x)-
\bar{\psi} (x)P_rt^a\frac{\delta S}{\delta\bar{\psi}(x)}\right]
-\partial_\mu\left[\frac{\delta S}{\delta {A^a}_\mu(x)}\right]
+ief^{abc}{A^b}_\mu \frac{\delta S}{\delta{A^c}_\mu}=0
\end{displaymath}
In Hamiltonian formulation the invariance (1.4) should appear as a
set of the first-class constraints which generates gauge transformations
in the phase space.
Weakly conserving right current corresponds to the global symmetry
transformations of the form
\begin{displaymath}
\delta_\alpha\psi=i\alpha P_r\psi, \quad
\delta_\alpha\bar\psi =-i\alpha P_l\bar\psi,
\end{displaymath}
\begin{equation}
\delta_\alpha A_\mu=0, \quad \alpha=\alpha^at^a=const.
\end{equation}
The conservation law of the right current becomes strong when the spinor
connection (1.3) is changed by another one which involves the additional
pseudoscalar fields $\phi = {\phi}^a(x)t^a$ in the following form:
\begin{eqnarray}
\Gamma_\mu=eA_\mu P_l+K_\mu P_r, \quad
{\bar{\Gamma}}_\mu=eA_\mu P_r+K_\mu P_l,\\
K_\mu(x)={K^a}_\mu(x)t^a, \quad
{K^a}_\mu=\partial_\mu\phi^b(x)E^{-1\,ba}(x).\nonumber\\
E^{-1\,ba}(x)={\left[\frac{1-\exp\left(-V(x)\right)}{V(x)}\right]}^{ba},
\quad
V^{ab}=-i\phi^cf^{abc}\nonumber
\end{eqnarray}
The theory (1.1) with the connection (1.6) possesses both the invariance
(1.4) and the new symmetry which is gauged by the fields $\phi^a (x)$:
\begin{eqnarray}
& &\delta_\alpha\psi=i\alpha(x)P_r\psi, \quad
\delta_\alpha\bar\psi =-i\alpha(x)\bar\psi P_l,\nonumber\\
& &\delta_\alpha\phi^a=\alpha^a(x) E^{ba}, \quad \delta_\alpha A_\mu=0,\\
& &E^{ba}(x)={\left[\frac{V(x)}{1-\exp\left(-V(x)\right)}\right]}^{ba}.
\nonumber
\end{eqnarray}
When the auxiliary fields are introduced in the action, than the new
first-class constraints appear in the Hamiltonian formalism to generate
the local transformations (1.7) on the theory phase space.  The connection
(1.6) apparently coincides with the standard one (1.3) being the gauge
condition $\phi^a=0$ imposed, that explicitly confirms classical
equivalence of the both formulations of the nonabelian chiral theory. On
the other hand the formulation using a localized invariance (that is the
connection (1.7)) allows to regard the both symmetries on equal footing
from the standpoint of the BFV-quantization method.  In this case,
particularly, the BRST-charge will contain the first-class constraints
corresponding to both of the invariances.  Thus the anomalies of the full
set theory symmetries will be described in a unified form by the unique
set of the gauge algebra generating equations. We will study in this paper
the quantum theory based on these generating equations.

Anomaly free operator canonical quantization of the theory (1.1) within
the BFV--approach \cite{batvil,batfra,batfrad} implies resolving of the
following foremost problems:
\begin{quote}
i) it is necessary to define the consistent regularization of the
operator functionals (especially those ones which are actually involved
into the principal generating equations) including the products of
the noncommuting operators as well as the derivatives of such products
at the same space point. In so doing the
normal ordering should be introduced in a manner providing the regularized
the operator functional algebra to be well-defined in the Fock space.\\
ii) it implies to solve operatorial generating equations of the gauge
algebra \cite{batfra} --that is to find the nilpotent operator of the
BRST-charge $\Omega $ and the Hamiltonian $\cal H$ commuting with
$\Omega $. It is important to examine that the Schwinger terms in the
commutators of regularized physical operators do not break the BRST--
invariance.
\end{quote}
Both of the requirements are studied in the paper and the solution is
shown to be existed only for the first one within the given set of the
fields and invariances. The Schwinger terms in the commutators of the
generating operators $\Omega $ and $\cal H$ are evaluated explicitly,
they describe the BRST-anomaly and never vanish.

The paper is organized as follows. In section 2 the regularized
operator Hamiltonian formulation of the constraint theory (1.1) is
constructed. The regularization scheme is chosen in a manner providing the
explicit covariance for the operator functionals to be considered.
The Schwinger terms in
the commutators among the regularized first-class constraint operators
and the Hamiltonian are further evaluated. It should be pointed
out that it is the commutator of the constraints with the total
Hamiltonian rather then with the Hamiltonian part involving current
that make a contribution to the anomalous Schwinger terms of the
involution relations (that is the current algebra is not enough to describe
the anomalies of the constraint algebra).
In so doing the anomaly in the conservation conditions of the
constraints appears to be covariant and consistent with the covariant
divergence anomaly of the left current obtained by Fujikawa \cite{fuj,fuji}.
Note that no gauge for ${A^a}_\mu$ was used in the paper including the
evaluation of the commutators of the constraints with the Hamiltonian
and between them (in contrast to other papers on the Hamiltonian
approach to the current algebra). In section 3 the anomaly in the
Hamiltonian generating equations of the gauge algebra is evaluated.
The nilpotency and conservation of the BRST-charge was demonstrated to
be broken at the quantum level. In section 4 the axial invariance was
presented as gauge one by the abelian analog of the connection (1.6).
In much the same way as in sections 2 and 3 the axial anomaly is shown
to be transform to the BRST-anomaly. The conclusion contains the discussion
of the Wess--Zumino conditions for the anomaly and the concluding remarks.
In appendix A all the quantum commutators contributing to the generating
equations of the gauge algebra are evaluated. In appendix B all the vacuum
expectation values which are needed for quantum commutators of the
Hamiltonian nonabelian chiral and abelian axial theories are evaluated.
\setcounter{equation}{0}
\section{Constrained Hamiltonian formulation:\protect\\ regularization
and Schwinger terms}
Let us introduce in the theory (1.1) the momenta $p^a_k({\bf x},t)$
canonically conjugated to $A^{ak}({\bf x},t)$:
\begin{equation}
p^a_k({\bf x},t)=\frac{\partial{\cal L}}{\partial(\partial_0A^{ak}
({\bf x},t))}=F^a_{0k}({\bf x},t)
\end{equation}
whose equal time commutators are:
\begin{equation}
\left[ A^{ai}({\bf x},t),p^b_k({\bf y},t) \right]=i\delta^i_k\delta^{ab}
\delta ({\bf x}-{\bf y}).
\end{equation}
The canonical commutation relations for the fermions are of the form:
\begin{equation}
\left[\psi({\bf x},t),\bar{\psi}({\bf y},t)\right]=-\gamma^0\delta
({\bf x}-{\bf y})
\end{equation}
(the commutator is implied here for the bosonic variables and the
anticommutator -- for the fermionic ones)

There are the Gauss constraints in the Hamiltonian formulation of
the theory (1.1)
\begin{eqnarray}
G^a_l(x)&=&(\stackrel{l}{\nabla}_kp^k(x))^a+ej^{a0}_l(x),\\
{j_{\stackrel{{\scriptstyle{l}}}{{\scriptstyle{r}}}}}^{a\mu}&=
&\bar{\psi}\gamma^\mu t^aP_{\stackrel{{\scriptstyle{l}}}{{\scriptstyle{r}}}}
\psi,\nonumber\\
(\stackrel{l}{\nabla}_kp^k(x))^a&=&\partial_kp^{ak}(x)-ief^{abc}A^{bk}p^c_k,
\nonumber
\end{eqnarray}
which generate the gauge transformations (1.4) in the phase space. If the
invariance (1.5) is considered as a local one, then the connection
(1.1) incorporates the fields $\phi^a$\/(1.6), hence $\phi^a$ and
the momenta $p^a$ must be regarded as a one more canonical pair in the
theory phase space:
\begin{equation}
\left[ \phi^a({\bf x},t),p^b({\bf y},t)\right] =i\delta^{ab}\delta
({\bf x}-{\bf y}).
\end{equation}
Then the following constraints appear in the Hamiltonian formulation
\begin{equation}
G^a_r(x)=p^a(x)-E^{-1ab}(x)j^{b0}_r(x),
\end{equation}
to generate the local transformations (1.7) on the phase space.

The Hamiltonian of the theory is of the form:
\begin{eqnarray}
& &H=H_M+H_F,\ H_F(t)=-i\int^{}_{}d{\bf x}\bar{\psi}({\bf x},t)\gamma^k
\nabla_k\psi({\bf x},t),\\
& &H_M(t)=\int^{}_{}d{\bf x}\,\left\{\frac{1}{2}p^{a}_k({\bf x},t)p^{ak}
({\bf x},t)+\frac{1}{4}F^a_{kl}({\bf x},t)F^{a\,kl}({\bf x},t)\right\}.
\nonumber
\end{eqnarray}
On the classical level the constraints (2.4),(2.6) are nonabelian first
class constraints commuting with the Hamiltonian (2.7).

In quantum theory the expressions incorporating the products of noncommuting
operators at the same points as well as their derivatives need a
regularization. The problem of regularization in the operator Hamiltonian
formulation connected with the issue of normal ordering as far the
operators should be well defined in the Fock space. For the sake of
unitarity in the physical subspace, the regularization and normal
ordering must be compatible with the gauge invariance of the physical
operators. The regularization is proposed for the quadratic forms of
fermions to be consistent to the specific connection choice (as it is
illustrated below by the example of the current and the covariant
derivative thereof). The suggested regularization represents some
modification of well-known point-splitting regularization \cite{jack}.
The correct transformation properties of the product of the fermionic
operators with the space-separated arguments are provided with multiplying
by the specially constructed splitting function
$R_{\stackrel{{\scriptstyle{l}}}{{\scriptstyle{r}}}}^
{\mu a}({\bf x},t;\mbox{{\boldmath$\varepsilon$}})$, which is
represented as a power series in the regularization parameter up to the
third order inclusively, because the fermion propagator diverges as no
more than $\varepsilon^{-3}$. It is the first order in the
$\varepsilon^{-1}$ which defines the anomaly value, while the second and
the third orders do not contribute because of the regularizing function
${R_{\stackrel{{\scriptstyle{l}}}{{\scriptstyle{r}}}}}^{\mu a}({\bf x},t;
{\mbox{\boldmath$\varepsilon$}})$
is taken up to the third order.
\begin{eqnarray}
& &{j_{\stackrel{{\scriptstyle{l}}}{{\scriptstyle{r}}}} }^{\mu a\, reg}
({\bf x},t;{\mbox{\boldmath$\varepsilon$}})=\bar{\psi}({{\bf x}}_+,t)
{R_{\stackrel{{\scriptstyle{l}}}{{\scriptstyle{r}}}}}^{\mu a}({\bf x},t;
{\mbox{\boldmath$\varepsilon$}})%
\psi({\bf x}_-,t),{\bf x}_\pm={\bf x}\pm
\frac{{\mbox{\boldmath$\varepsilon$}}}{2}\\
& &{R_{\stackrel{{\scriptstyle{l}}}{{\scriptstyle{r}}}}}^{\mu a}
({\bf x},t;{\mbox{\boldmath$\varepsilon$}})={\bf e}^+({\bf x},t;
{\mbox{\boldmath$\varepsilon$}})\gamma^\mu t^a
P_{\stackrel{{\scriptstyle{l}}}{{\scriptstyle{r}}}}{\bf e^-}({\bf x},t;
{\mbox{\boldmath$\varepsilon$}})+\\
& &{}+\frac{1}{6}\gamma^\mu t^a
P_{\stackrel{{\scriptstyle{l}}}{{\scriptstyle{r}}}}{
\left[{\bf \bar{e}^+}({\bf x},t;{\mbox{\boldmath$\varepsilon$}}),
{\bf e^-}({\bf x},t;{\mbox{\boldmath$\varepsilon$}})\right]}_-,\nonumber\\
& &{\bf e^+}({\bf x},t;{\mbox{\boldmath$\varepsilon$}})=
\exp\left\{i\int^{{\bf x_+}}_{{\bf x}}\bar{\Gamma}_j({\bf z},t)\,
d{\bf z}^j\right\},\nonumber\\
& &{\bf e^-}({\bf x},t;{\mbox{\boldmath$\varepsilon$}})=
\exp\left\{i\int^{{\bf x}}_{{\bf x_-}}\Gamma_j({\bf z},t)\,
d{\bf z}^j\right\},\nonumber\\
& &{\bf \bar{e}^+}({\bf x},t;{\mbox{\boldmath$\varepsilon$}})=
{\bf e^+}({\bf x},t;{\mbox{\boldmath$\varepsilon$}})
\left(\bar{\Gamma}_j\to\Gamma_j\right).\nonumber
\end{eqnarray}
The local form of regularized currents and covariant derivatives thereof are
defined as:
\begin{eqnarray}
& &{j_{\stackrel{{\scriptstyle{l}}}{{\scriptstyle{r}}}}^{\mu a}({\bf x},t)
=\lim_{{\mbox{\boldmath$\varepsilon$}}\to 0}
j_{\stackrel{{\scriptstyle{l}}}{{\scriptstyle{r}}}}^{\mu a\,reg}({\bf x},t;
{\mbox{\boldmath$\varepsilon$}})},\\
& &\stackrel{r}{\nabla}_k
j_{\stackrel{{\scriptstyle{l}}}{{\scriptstyle{r}}}}^{ka}({\bf x},t)=
\lim_{{\mbox{\boldmath$\varepsilon$}}\to 0}\bigg\{\partial_k
j_{\stackrel{{\scriptstyle{l}}}{{\scriptstyle{r}}}}^{ka\,reg}
({\bf x},t;{\mbox{\boldmath$\varepsilon$}})-\nonumber\\
& & {}-if^{abc}K^b_k({\bf x},t)
j_{\stackrel{{\scriptstyle{l}}}{{\scriptstyle{r}}}}^{kc\,reg}
({\bf x},t;{\mbox{\boldmath$\varepsilon$}})\bigg\} ,\\
& &\stackrel{l}{\nabla}_k
j_{\stackrel{{\scriptstyle{l}}}{{\scriptstyle{r}}}}^{ka}({\bf x},t)=
\lim_{{\mbox{\boldmath$\varepsilon$}}\to 0}\bigg\{ \partial_k
j_{\stackrel{{\scriptstyle{l}}}{{\scriptstyle{r}}}}^{ka\,reg}({\bf x},t;%
{\mbox{\boldmath$\varepsilon$}})-ief^{abc}A^b_k({\bf x},t)
j_{\stackrel{{\scriptstyle{l}}}{{\scriptstyle{r}}}}^{kc\,reg}({\bf x},t;
{\mbox{\boldmath$\varepsilon$}})\bigg\}.\nonumber
\end{eqnarray}
The derivatives and commutators are defined hereby in the following sense:
at first all the differentiation and commutation of corresponding
regularized operators perform and then the limit
${\mbox{\boldmath$\varepsilon$}}\to 0$ is taken as is done for example for a
partial derivative in (2.11). The limit is taken to provide the spatial
isotrophy property:
\begin{equation}
\lim_{{\mbox{\boldmath$\varepsilon$}}\to 0}\left(
\frac{\mbox{{\boldmath$\varepsilon^i$}}\mbox{{\boldmath$\varepsilon^j
$}}}{{\mbox{\boldmath$\varepsilon$}}^2}\right)=\frac{1}{3}\delta^{ij}.
\end{equation}
If one need to define the quantum corrections to equal-time commutators
of the fermion quadratic forms, in particular, of the currents, constraints
and Hamiltonian, it will be sufficient to know an equal-time normal coupling
of fermions. It is evident from Wick theorem that the Schwinger terms are
defined thereby as the contributions with two couplings. Let us introduce the
fermionic vacuum $\langle0\vert$ in the given fields $A^\mu$ and $\phi^a$
(the last mentioned---for the case of connection (1.6) only). Let us define
the normal ordering of fermions in respect to the given vacuum in such a way
that the normal coupling is the equal-time limit of the fermion propagator.
\begin{eqnarray}
& &\bar{\psi}({\bf x_+},t)\psi({\bf x_-},t)=i\lim_{\varepsilon^0\to-0}
S^c(x_-,x_+)=iS^c({\bf x_-},t;{\bf x_+},t),\\
& &\hspace{1mm}\rule{0.4pt}{2mm}\rule{3.5em}{0.4pt}%
\rule{0.4pt}{2mm}\nonumber
\end{eqnarray}
where $\varepsilon=(\varepsilon^0,{\mbox{\boldmath$\varepsilon$}}),
S^c(x_-,x_+)$  is the propagator
\begin{displaymath}
S^c(x_-,x_+)=-i\langle 0\vert T\bar{\psi}(x_+)\psi(x_-) \vert 0 \rangle,
\end{displaymath}
obeying the equation
\begin{equation}
i\gamma^\mu\nabla_\mu^{(x)}S^c(x,y)=-\delta(x,y)
\end{equation}
with the corresponding connection in the covariant derivative.

In the case of this normal ordering the Schwinger terms in the considered
commutators can be expressed in terms of the corresponding vacuum
expectation values of the quadratic forms of the fermions. The averaging
is taken with respect to the fermionic vacuum in given fields $A_\mu$ and
$\phi$. All the vacuum expectation values are obtained in appendix B by
means of the Schwinger--de Witt method \cite{dew}. Let us at first evaluate
the divergence of the left and right currents in the chiral nonabelian
theory without additional fields (the connection (1.2)) using the
regularized quantum commutators, which are calculated in appendix A, and
then the quantum constraint algebra of the same theory with the additional
fields (the connection (1.6)) is examined too.

By making use of the Heisenberg equations with the Hamiltonian (2.7) without
additional fields ($\phi=0$) and the vacuum expectation values
of the commutators
$C_{\stackrel{{\scriptstyle{l}}}{{\scriptstyle{r}}}}^3(g;t)$,
$C_{\stackrel{{\scriptstyle{l}}}{{\scriptstyle{r}}}}^4(g;t)$
(A.4;A.8;A.9) we obtain
\begin{eqnarray}
& &\langle 0\vert\stackrel{l}{\nabla}_\mu j^\mu_l(g;t)\vert 0\rangle=\\
& &=\langle 0\vert\stackrel{l}{\nabla}_ij^i_l(g;t)\vert 0\rangle+
\langle 0\vert i\left[H(t),j^0_l(g;t)\right]\vert 0\rangle=\nonumber\\
& &\mbox{}=\frac{e^2}{32\pi^2}\int^{}_{}d{\bf y}Sp\left(g({\bf y},t)
\tilde{F}^{\alpha\beta}({\bf y},t)F_{\alpha\beta}({\bf y},t)\right),
\nonumber\\
& &\langle 0\vert\stackrel{r}{\nabla}_\mu j^\mu_r(g;t)\vert 0\rangle=
\langle 0\vert\partial_ij^i_r(g;t)\vert 0\rangle+\nonumber\\
& &+\langle 0\vert i\left[H(t),j^0_r(g;t)\right]\vert 0\rangle=0,\nonumber
\end{eqnarray}
where J(f) means
\begin{displaymath}
J(f)=\int^{}_{}d{\bf x}\,f^a({\bf x})J^a({\bf x}).
\end{displaymath}

At the classical level the both of these divergences are obviously vanished,
while the quantum divergence of the left current contains nonvanishing
anomaly (2.15). In the constrained Hamiltonian formulation the
corresponding anomaly arises in the involution relation of the Gauss law
with the Hamiltonian as it is shown below. The anomaly in the left current
divergence derived here turns out to be explicitly covariant (although
some vacuum expectation values may be noncovariant, for example see (A.7))
and it concurs with the covariant anomaly described by the path integral
method \cite{fuj,fuji} but differs from the results of some other operator
calculations \cite{ghoban,seo}.

Let us consider the theory (1.1) with the Hamiltonian (2.7), connection
(1.6) and the constraints (2.4), (2.6). Applying the quantum commutators
from appendix A we obtain the following quantum corrections to the
constraint algebra:
\begin{eqnarray}
& &\langle0\vert\left[G_l(g,t),G_r(f,t)\right]\vert0\rangle=0,\\
& &\langle0\vert\left[G_r(g,t),G_r(f,t)\right]\vert0\rangle=0,\\
& &\langle0\vert i\left[H(t),G_r(f,t)\right]\vert0\rangle=0,\\
& &\langle0\vert\left[G_l(g,t),G_l(f,t)\right]\vert0\rangle=
C^8(g,f;t)+e^2C^{20}_l(g,f;t)+\\
& &\mbox{}+eC^1_l(g,f;t)-eC^1_l(f,g;t)=e\int^{}_{}d{\bf y}f^{abc}g^a
({\bf y},t)f^b({\bf y},t)G^c_l({\bf x},t)+\nonumber\\
& &\mbox{}+\frac{ie^3}{48\pi^2}\int^{}_{}d{\bf y}\tilde{F}^{a0i}({\bf y},t)
\left(\stackrel{l}{\nabla}_ig({\bf y},t)\right)^bf^c({\bf y},t)d^{abc},
\nonumber
\end{eqnarray}
where $d^{abc}=Sp\left(t^a{\left[t^b,t^c\right]}_+\right)$.
\begin{eqnarray}
& &\langle 0\vert i\left[H(t),G_l(f,t)\right]\vert 0\rangle=
C^6(g,t)+eC^3_l(g,t)+\\
& &+eC^4_l(g,t)=\frac{e^3}{32\pi^2}\int^{}_{}d{\bf y}\,Sp\left(g({\bf y},t)
\tilde{F}^{\alpha\beta}({\bf y},t)F_{\alpha\beta}({\bf y},t)\right).
\nonumber
\end{eqnarray}

The Schwinger terms in the Gauss law, that are seen from (2.16-2.20),
may indicate the possible breaking of the gauge invariance (1.4)
at the quantum level. This result does not depend on the presence of the
auxiliary fields. The gauge invariance connected with the additional fields
$\phi$ do not induce anomalous Schwinger terms. The quantum corrections to
the commutator (2.20) are explicitly covariant and they correspond to the
anomaly in the left current divergence (2.15). The anomaly in Gauss law
commutators (2.19) is covariant and it differs from the consistent
anomaly satisfied Wess-Zumino conditions \cite{weszum} which has been
considered in \cite{fadsha}--\cite{fujik}. The covariant anomaly in Gauss
law commutators found in \cite{ghoban,hosseo} differs from (2.19) by the
coefficient. The anomaly in Gauss law commutators with the Hamiltonian
(2.20) as well as in the left current divergence (2.15) derived by the
Heisenberg equation was considered for example in \cite{seo} with the
result consistent to the Wess-Zumino conditions and in \cite{hosseo} with
the result agreed with our one. It should be mentioned, however, that
we do not use gauge fixing conditions in our consideration in contrast
to \cite{ghoban,hosseo}.

The mere fact that there are Schwinger terms in constraint algebra does not
imply an anomaly appearance, it may  also be related to usual deformation of
the gauge group representation. For example in the string theory the central
extension of the Virasoro algebra does not imply an anomaly. However the
loss of the nilpotency of the BRST charge and (or) the breaking of its
conservation at a sacrifice of the Schwinger terms should be actually
treated as a true manifestation of the anomaly.
In the following section the question is studied how the generating
equations are affected by the Schwinger terms.
\setcounter{equation}{0}
\section{The chiral anomaly and the BFV generating equations}
In accordance to the general prescription of the BFV method let us introduce
the ghosts
${\cal C}^a_l(x)$, ${\cal C}^a_r(x)$, ${\cal P}^a_l(x)$, ${\cal P}^a_r(x)$,
$\bar{\cal P}^a_l(x)$, $\bar{\cal P}^a_r(x)$, $\bar{\cal C}^a_l(x)$,
$\bar{\cal C}^a_r(x)$; the Lagrange multipliers
$\lambda^a_l$, $\lambda^a_r$, $\pi^a_l$, $\pi^a_r$ which have the following
statistics, ghost numbers  and (anti)commutation relations:
\begin{eqnarray*}
& &gh({\cal C}^a_l)=1,\  gh({\cal C}^a_r)=1,\  gh({\cal P}^a_l)=1,\
gh({\cal P}^a_r)=1,\\
& &gh(\bar{\cal P}^a_l)=-1,\  gh(\bar{\cal P}^a_r)=-1,\
gh(\bar{\cal C}^a_l)=-1,\  gh(\bar{\cal C}^a_r)=-1,\\
& &gh(\lambda^a_l)=0,\  gh(\lambda^a_r)=0,\  gh(\pi^a_l)=0,\
gh(\pi^a_r)=0,
\end{eqnarray*}
\begin{eqnarray*}
\left[{\cal C}^a_l({\bf x},t),\bar{\cal P}^b_l({\bf y},t)\right]=
i\delta^{ab}\delta({\bf x}-{\bf y})&,&\left[{\cal C}^a_r({\bf x},t),
\bar{\cal P}^b_r({\bf y},t)\right]=i\delta^{ab}\delta({\bf x}-{\bf y}),\\
\left[{\cal P}^a_l({\bf x},t),\bar{\cal C}^b_l({\bf y},t)\right]=
i\delta^{ab}\delta({\bf x}-{\bf y})&,&\left[{\cal P}^a_r({\bf x},t),
\bar{\cal C}^b_r({\bf y},t)\right]=i\delta^{ab}\delta({\bf x}-{\bf y}),\\
\left[\pi^a_l({\bf x},t),\lambda^b_l({\bf y},t)\right]=
i\delta^{ab}\delta({\bf x}-{\bf y})&,&\left[\pi^a_r({\bf x},t),
\lambda^b_r({\bf y},t)\right]=i\delta^{ab}\delta({\bf x}-{\bf y}).
\end{eqnarray*}
The other possible (anti)commutators vanish.The Lagrange multipliers
$\lambda^a_l$,$\lambda^a_r$, $\pi^a_l$,$\pi^a_r$ are bosons, the ghosts
${\cal C}^a_l(x)$,\ ${\cal C}^a_r(x)$, ${\cal P}^a_l(x)$,\
${\cal P}^a_r(x)$, $\bar{{\cal P}}^a_l(x)$, $\bar{{\cal P}}^a_r(x)$,
$\bar{{\cal C}}^a_l(x)$, $\bar{\cal C}^a_r(x)$ are fermions.
Let regard the fermionic and bosonic generating operators of the gauge
algebra corresponding to the constraints (2.4),(2.6) and Hamiltonian (2.7)
(with the connection (1.6)).
\begin{eqnarray}
& &\Omega_{min}=\int^{}_{}d{\bf x}\bigg({\cal C}^a_l({\bf x},t)
G^a_l({\bf x},t)+{\cal C}^a_r({\bf x},t)G^a_r({\bf x},t)+\nonumber\\
& &{}+\frac{ie}{2}f^{abc}{\cal C}^a_l({\bf x},t){\cal C}^b_l({\bf x},t)
\bar{\cal P}^c_l({\bf x},t)\bigg),\nonumber\\
& &\Omega=\Omega_{min}+\int^{}_{}d{\bf x}\,\left(\pi^a_l({\bf x},t)
{\cal P}^a_l({\bf x},t)+\pi^a_r({\bf x},t){\cal P}^a_r({\bf x},t)\right),\\
& &{\cal H}=H.
\end{eqnarray}
The total unitarizing Hamiltonian is of the form:
\begin{equation}
H_\psi={\cal H}-i\left[\Psi(t),\Omega(t)\right]
\end{equation}
where $\Psi$ is the gauge fixing fermion:
\begin{displaymath}
\Psi=\int^{}_{}d{\bf x}\left(\bar{{\cal C}}^a_l\chi^a_l+\bar{{\cal C}}^a_r
\chi^a_r+\lambda^a_l\bar{{\cal P}}^a_l+\lambda^a_r\bar{{\cal P}}^a_r\right).
\end{displaymath}
$\chi^a_l$ and  $\chi^a_r$  are the spatial parts of the relativistic gauges
$\Phi^a_l$, $\Phi^a_r$:\vspace{2mm} $\Phi^a_l=\dot{\lambda^a_l}-\chi^a_l$,
$\Phi^a_r=\dot{\lambda^a_r}-\chi^a_r$.\vspace{2mm}
As $\Phi^a_l$ it can be chosen the Lorentz gauge ($\lambda=A^0$) and
$\Phi^a_r=\Box\phi^a$ (because of $\lambda^a_r=\dot{\phi^a}$).
If the theory is anomaly free, the BRST charge (3.1) must be nilpotent
and commute with the Hamiltonian. By using the constraint algebra
found in previous section as well as the involution relations of the
constraints and Hamiltonian the following (anti)commutators are
derived:
\begin{eqnarray}
\left[\Omega_{min},\Omega_{min} \right] &=&\frac{ie^3}{48\pi^2}\int^{}_{}
d{\bf y}\,\tilde{F}^{a0i}({\bf y},t){\left( \stackrel{l}{\nabla}_i
{\cal C}_l({\bf y},t)\right)}^b{\cal C}_l^c({\bf y},t)d^{abc},\\
i\left[{\cal H},\Omega_{min}\right]&=&\frac{e^3}{32\pi^2}\int^{}_{}
d{\bf y}\,Sp\left({\cal C}_l({\bf y},t)\tilde{F}^{\alpha\beta}({\bf y},t)
F_{\alpha\beta}({\bf y},t)\right).
\end{eqnarray}

It is essential that the commutators of the terms cubic in ghosts do not
have Schwinger terms. It is the circumstance that distinguishes the
four-dimensional chiral symmetry from the two dimensional conformal one
from the standpoint of the BFV formalism. In the $d=2$ case the ghost
Schwinger terms are not trivial and they may cancel the contribution of the
Schwinger terms to the constraints commutators.

The relations (3.4) and (3.5) show that the both quantum generating
equations
do not valid for the generating operators (3.1),(3.2). The loss of the
nilpotency of $\Omega$ and the breakdown ${\cal H}$ BRST invariance in
the quantum theory means the chiral BRST anomaly in the true sense. It can
be also shown that there are no local counterterms which could bring both
the anomalous commutators (3.4), (3.5) to zero if they were added to the
Hamiltonian and constraints. Thus there are no ways to remove the chiral
anomaly from the generating equations. Although in the
framework of the given model (1.1) with the invariances (1.4), (1.7) the
chiral anomaly can not be eliminated, the previous consideration shows
that the ghost sector may also have nontrivial Schwinger terms in the
theory with a more wide invariance generated by first class constraints
which have to be chiral spinors (and by the original constraints too).
If the Gauss constraint might have nontrivial involution relations
with the chiral spinor constraints, the cancelation of the anomalous
chiral contribution of the form (3.4),(3.5) may become possible
due to the Schwinger terms of the spinor ghost commutators.
However, we can not find now an appropriate model of the desirable first
class constraint structure to provide the mutual cancelation of the
anomalous contributions to the generating equations.
\setcounter{equation}{0}
\section{Axial anomaly}
Axial anomaly or Adler-Bell-Jackiw one \cite{adl,beljac} historically was
one of the first obtained and for many years it has been serving as a
touchstone for new methods as well as for polishing traditional approaches
to description an anomaly. In given section the techniques for the
transformation of the axial anomaly to the corresponding BRST anomaly
is suggested which allows to use the universal means of the Hamiltonian
BFV-BRST formulation for description this anomaly. The Schwinger terms
are evaluated
much as it was done for the chiral anomaly in sections 2, 3 and appendixes.
Therefore only final results are written out here -- the anomaly in the
conservation law of the axial current and the anomaly in the generating
equations of the gauge algebra.

The theory of massless fermions interacting with the massless abelian
vector field is described by the action (1.1) in the abelian case.
If the fermion connection in the covariant derivative (1.2) includes the
vector field $\Gamma_\mu=$ ${}=\bar{\Gamma}_\mu=eA_\mu$ only, the theory
possesses an abelian gauge symmetry and a global axial one. Corresponding
transformations are of the form:
\begin{eqnarray}
\delta_\alpha\psi=ie\alpha\psi&,&\delta_\alpha\bar{\psi}=-ie\alpha
\bar{\psi}\; ,\;  \delta_\alpha A_\mu=\partial_\mu\alpha;\\
\delta_{\alpha_5}\psi=i\alpha_5\gamma^5\psi&,&\delta_{\alpha_5}
\bar{\psi}=-i\alpha_5\bar{\psi}\gamma^5\; ,\; \delta_{\alpha_5}A_\mu=0,
\end{eqnarray}
where $\alpha_5=const,\alpha=\alpha(x)$. Classically the invariance
(4.1) leads to the identity connected to the strong charge conservation --
the Gauss law. In the Hamiltonian formulation Gauss law is a first class
constraint generating the gauge transformation on the phase space. The
invariance (4.2) corresponds to the weakly conserved gauge invariant
current. Because of this, in the Hamiltonian formulation the corresponding
charge must commute with the Hamiltonian as well as with the first class
constraint.

The conservation law of the axial current becomes strong, if the gradient of
a pseudoscalar field $\phi_5$ being treated as a new independent variable of
the theory (1.1) is introduced into the spinor connection:
\begin{equation}
\Gamma_\mu=eA_\mu-\gamma^5\partial_\mu\phi_5 ,\; \bar{\Gamma}_\mu=
eA_\mu+\gamma^5\partial_\mu\phi_5.
\end{equation}
Then the spinor transformations (4.2) become local and the field $\phi_5$
becomes pure gauge field:
\begin{equation}
\alpha_5=\alpha_5(x)\;,\;\phi^{(\alpha_5)}_5=\phi_5-\alpha_5.
\end{equation}
In so doing the first class constraint appears to generate local axial
transformations (4.2),(4.4) on the phase space. The theory with the
connection (4.3) coincides to the standard one under the gauge condition
$\phi_5=0$. Thus the classical equivalence of the both theories
is obvious. At the same time the formulation using localized axial
invariance (the connection (4.3)) allows to regard the both symmetries on
equal footing. In particular, in that case the BRST charge includes the
first class constraints corresponding to both invariances. By this means
the anomalies of the symmetries (4.1), (4.2), (4.4) are described by the
same generating equations of the gauge algebra.  The standard canonical
relations (2.2), (2.3) are rewritten for the abelian case. The gauge field
$\phi_5$ and the momenta $p_5$ conjugated to them are regarded as a
canonical pair:
\begin{displaymath} \left[\phi_5({\bf x},t),p_5({\bf
y},t)\right]=i\delta({\bf x}-{\bf y}) \end{displaymath}
Then there is the Gauss constraint in the theory (1.1):
\begin{equation}
T({\bf x},t)=\partial_kp^k({\bf x},t)+ej^0({\bf x},t)
\end{equation}
generating gauge transformation (4.1) in the phase space. If the axial
invariance is considered as a local one, that is the
connection in (1.1) and (2.7) includes the field $\phi_5$, then there is
one more constraint in the Hamiltonian formulation
\begin{equation}
T_5({\bf x},t)=p_5({\bf x},t)-j^0_5({\bf x},t)
\end{equation}
which generates local axial transformations (4.2), (4.4). The Hamiltonian
of the theory (1.1) is of the form (2.7) with the connection (4.3) in the
abelian case. The constraints (4.5), (4.6) are of the first class and they
are classically commuting with the Hamiltonian.

The following notations are used here:
\begin{displaymath}
j^\mu_{(5)}({\bf x},t)=(j^\mu({\bf x},t),j^\mu_5({\bf x},t))=
\bar{\psi}({\bf x},t)(\gamma^5)\gamma^\mu\psi({\bf x},t)
\end{displaymath}

The following regularization accounting the connection is suggested for
quadratic forms of fermions at the same point:
\begin{eqnarray*}
& &j^{\mu reg}_{(5)}({\bf x},t;{\mbox{\boldmath$\varepsilon$}})=
\bar{\psi}({\bf x_+},t)R^\mu_{(5)}
({\bf x},t;{\mbox{\boldmath$\varepsilon$}})\psi({\bf x_-},t) ,\;
{\bf x_\pm}={\bf x}\pm\frac{\mbox{\boldmath$\varepsilon$}}{2},\\
& &R^\mu_{(5)}({\bf x},t;\mbox{\boldmath$\varepsilon$})=
(\gamma^5)\gamma^\mu\exp\left(i\int^{{\bf x_+}}_{{\bf x_-}}
\bar{\Gamma}_j({\bf z},t)\,d{\bf z}^j\right).
\end{eqnarray*}
The local form of regularized currents and derivatives is defined in much
the same way as in (2.10) and (2.11).

By using canonical commutation relations only as well as the vacuum
expectation values $V^\nu_{(5)}({\bf x},t;{\mbox{\boldmath$\varepsilon$}})$
(B.14) and the space isotrophic limit (2.12) we are coming to the following
values of the quantum commutators:
\begin{eqnarray}
& &\langle 0\vert \left[ T(g,t),T(f,t) \right] \vert 0 \rangle=0\\
& &\langle 0\vert i\left[H(t),T(f,t)\right]\vert 0\rangle=0\\
& &\langle 0\vert\left[T_5(g,t),T_5(f,t)\right]\vert 0\rangle=0,\\
& &\langle 0\vert\left[T(g,t),T_5(f,t)\right]\vert 0\rangle=\\
& &\mbox{}=\frac{ie^2}{12\pi^2}\int^{}_{}d{\bf y}\,\partial_i
g({\bf y})f({\bf y})\tilde{F}^{i0}({\bf y},t),\nonumber\\
& &\langle 0\vert i\left[H(t),T_5(f,t)\right]\vert 0\rangle=\\
& &\mbox{}=-\frac{e^2}{4\pi^2}\int^{}_{}d{\bf x}\,\tilde{F}^{0i}
({\bf x},t)p_i({\bf x},t)f({\bf x},t)=\nonumber\\
& &\mbox{}=-\frac{e^2}{16\pi^2}\int^{}_{}d{\bf x}\,\tilde{F}^{\alpha\beta}
({\bf x},t)F_{\alpha\beta}({\bf x},t)f({\bf x},t).\nonumber
\end{eqnarray}

Thus, as one can see from (4.7--4.11), the axial anomaly of the theory
(1.1) with the connection (4.3) represents a quantum deformation of the
constraint algebra that allows to describe it in the framework of BFV
method in the same manner as it is done in the section 3 for the chiral
anomaly.

If the theory considered without the auxiliary gauge field $\phi_5$, the
quantum constraint algebra and the involution relation are of the form
(4.7),(4.8). Then the anomaly in the Gauss law in the constrained
Hamiltonian formulation is absent, while the weak conservation of the axial
current $j^\mu_5(f,t)$ and the axial invariance are broken in the quantum
theory:
\begin{eqnarray}
& &\partial_\mu j^\mu_5(f,t)=\langle 0\vert \partial_ij^i_5\vert 0\rangle+
\langle 0\vert i\left[H(t),j^0_5(f,t)\right]\vert 0\rangle=\\
& &\mbox{}=\frac{e^2}{4\pi^2}\int^{}_{}d{\bf x}\,\tilde{F}^{0i}({\bf x},t)
p_i({\bf x},t)f({\bf x},t)=\frac{e^2}{16\pi^2}\int^{}_{}d{\bf x}\,
\tilde{F}^{\alpha\beta}({\bf x},t)F_{\alpha\beta}({\bf x},t)f
({\bf x},t).\nonumber
\end{eqnarray}
It is apparent that in the theory without the additional field the
generating equations of the gauge algebra do not describe the anomaly.
 From this it follows that only the technique using additional field
$\phi_5$ allows to describe axial anomaly by the standard means of
Hamiltonian BFV-BRST method.

According to the general prescription of BFV method we introduce the ghosts
${\cal C}(x)$, ${\cal C}_5(x)$, ${\cal P}(x)$,${\cal P}_5(x)$,
$\bar{\cal P}(x)$, $\bar{\cal P}_5(x)$, $\bar{\cal C}(x)$,
$\bar{\cal C}_5(x)$;
the Lagrange multipliers $\lambda$,$\lambda_5$,$\pi$,$\pi_5$ with the
following statistics, ghost numbers and (anti)commutation relations:
\begin{eqnarray*}
gh({\cal C})=1,gh({\cal C}_5)=1 &,& gh({\cal P})=1, gh({\cal P}_5)=1,\\
gh(\bar{\cal P})=-1,gh(\bar{\cal P}_5)=-1 &,& gh(\bar{\cal C})=-1,
gh(\bar{\cal C}_5)=-1,\\
gh(\lambda)=0,gh(\lambda_5)=0 &,& gh(\pi)=0, gh(\pi_5)=0,\\
\left[{\cal C}({\bf x},t),\bar{\cal P}({\bf y},t)\right]=
i\delta({\bf x}-{\bf y})&,&\left[{\cal C}_5({\bf x},t),
\bar{\cal P}_5({\bf y},t)\right]=i\delta({\bf x}-{\bf y}),\\
\left[{\cal P}({\bf x},t),\bar{\cal C}({\bf y},t)\right]=
i\delta({\bf x}-{\bf y})&,&\left[{\cal P}_5({\bf x},t),
\bar{\cal C}_5({\bf y},t)\right]=i\delta({\bf x}-{\bf y}),\\
\left[ \pi({\bf x},t), \lambda({\bf y},t) \right] =
i \delta({\bf x}-{\bf y}) &,& \left[\pi_5 ({\bf x},t),
\lambda_5 ({\bf y},t)\right] =i\delta({\bf x}-{\bf y}).
\end{eqnarray*}
The other possible (anti)commutators vanish. The Lagrange multipliers
$\lambda$, $\lambda_5$, $\pi$, $\pi_5$ have the same statistics as the
constraints
$T$ (4.5), $T_5$ (4.6),the ghosts ${\cal C}(x)$, ${\cal C}_5(x)$,
${\cal P}(x)$, ${\cal P}_5(x)$,$\bar{\cal P}(x)$, $\bar{\cal P}_5(x)$,
$\bar{\cal C}(x)$, $\bar{\cal C}_5(x)$ have an opposite statistics to the
constraints. Let us regard the fermion and boson generating operators of
the gauge algebra corresponding to the constraints (4.5),(4.6) and the
Hamiltonian (2.7) (with the connection (4.3)).
\begin{eqnarray}
& &\Omega_{min}=\int^{}_{}d{\bf x}\left({\cal C}T+{\cal C}_5T_5\right),
\Omega=\Omega_{min}+\int^{}_{}d{\bf x}\left(\pi{\cal P}+\pi_5
{\cal P}_5\right)\\
& &{\cal H}=H
\end{eqnarray}
The total unitarizing Hamiltonian is of the form:
\begin{equation}
H_\psi={\cal H}-i\left[\Psi(t),\Omega(t)\right]
\end{equation}
$\Psi$ -- is the gauge fermion:
\begin{displaymath}
\Psi=\int^{}_{}d{\bf x}\left(\bar{\cal C}\chi+\bar{{\cal C}}_5\chi_5+
\lambda\bar{\cal P}+\lambda_5\bar{{\cal P}}_5\right)
\end{displaymath}
$\chi$ and $\chi_5$ are the space parts of the relativistic gauges.
The BRST charge (4.13) must be nilpotent and commuting with the Hamiltonian,
if there are no anomalies in the theory. By making use of the found above
constraint commutators and involution relations of the constraints with the
Hamiltonian (4.7--4.11) we obtain the following quantum (anti)commutators
for the charge and the Hamiltonian:
\begin{eqnarray}
& &\left[\Omega_{min},\Omega_{min}\right]=\frac{ie^2}{6\pi^2}
\int^{}_{}d{\bf x}\,{\cal C}\partial_j{\cal C}_5\tilde{F}^{0i},\\
& &i\left[{\cal H},\Omega_{min}\right]=-\frac{e^2}{4\pi^2}
\int^{}_{}d{\bf x}\,{\cal C}_5\tilde{F}^{0i}p_i=-\frac{e^2}{16\pi^2}
\int^{}_{}d{\bf x}\,{\cal C}_5\tilde{F}^{\alpha\beta}F_{\alpha\beta}.
\end{eqnarray}
The relations (4.16), (4.17) show that the nilpotency and the conservation
condition of the BRST charge is broken on the quantum level, which is a
consequence of an anomaly of the theory. The r.h.s. of the rel.(4.16),
(4.17) are just the BRST-anomaly induced by the quantum breakdown of an
axial symmetry of the theory and one can see the covariant character of this
anomalous contribution. In the next Section we discuss the Wess-Zumino
consistency conditions for the BRST-anomaly and summing up the results.
\setcounter{equation}{0}
\section{Discussion and Concluding Remarks}
The theory of massless fermions chiraly coupled to the nonabelian vector
field  is considered in the given paper as well as the axial abelian theory.
The anomalies in the operator BFV generating equations of the gauge
algebra are found in a covariant form. These anomalies are induced in
their turn by the anomalous Schwinger terms in the commutators of Gauss
law constraints with the Hamiltonian and between themselves. By making use
of the Heisenberg equations the anomaly in left current divergence is
obtained in the form which agreed with the covariant anomaly found before
by the path integral method. To keep the gauge invariant form in the
calculations we use the splitting point regularization technique with the
modified phase factor for the expressions quadratic in fermions and the
Schwinger-deWitt approach to find the Green function. The theory of
massless fermions interacting with the massless abelian vector field is
also considered in the paper: the anomaly is found in the generating
equations as well as in the divergence of axial current. In both of the
theories additional fields are introduced to treat the global as well
as the local invariances of the theory by means of the unique BRST charge
on an equal footing. The anomaly of the generating equations in the axial
theory arises by introducing additional field only which means that it
is impossible to provide the local and global symmetries simultaneously in
the quantum theory. While in the chiral theory the additional fields are not
essential for the quantum generating equations because it is the purely
gauge symmetry which is anomalous.

Let us discuss now the Wess--Zumino consistency conditions for the
covariant anomalies in the BFV--generating equations considered in the paper.
By making use of the rel (3.4),(3.5),(4.16),(4.17) for the BRST--anomaly we
may have the following expressions for the cycled double commutators of the
generating operators:
\begin{eqnarray}
&&J_1=\left[\Omega,\left[\Omega,\Omega\right]_+\right]_-=
-\frac{ie^4}{48\pi^2}\int^{}_{}d{\bf x}\,d^{abh}f^{hcd}\tilde{F}^{a0i}
\left(\nabla_i {\cal C}\right)^b{\cal C}^c{\cal C}^d\\
&&J_2=2\left[\Omega,i\left[H,\Omega\right]_-\right]_+-i\left[H,
\left[\Omega,\Omega\right]_+\right]_-=\frac{e^4}{24\pi^2}
\int^{}_{}d{\bf x}\,d^{abh}f^{hcd}\tilde{F}^{a0i}p^b_i{\cal C}^c
{\cal C}^d+\nonumber\\
&&{}+\frac{ie^3}{24\pi^2}\int^{}_{}d{\bf x}\,d^{abc}\varepsilon^{0inm}
p^a_i\left(\nabla_n {\cal C}\right)^b\left(\nabla_m {\cal C}\right)^c
\end{eqnarray}
For the axial anomaly one has for these commutators the following
expressions:
\begin{eqnarray}
&&J_1=0,\\
&&J_2=-\frac{ie^2}{3\pi^2}\int^{}_{}d{\bf x}\,\varepsilon^{0inm}
\partial_n p_m {\cal C}\partial_i {\cal C}_5
\end{eqnarray}
If the operator algebra in the anomalous theory was consistent, than
the r.h.s. of the rel.(5.1)--(5.4) would have to vanish to provide
the Jacoby identity for the double commutators in the l.h.s. As we have
already mentioned the covariant anomaly usually turns out to be
inconsistent with the Wess--Zumino condition and the relations
(5.1)--(5.4) demonstrate that the covariant BRST--anomaly is not an
exceptional one.

We would like to mention that it may sometimes seem to
be possible to transform a covariant anomaly into a consistent one by making
use of two different tools. The first is to change the commutation relations
between the momenta conjugated to the gauge fields

$$
\left[p^i,p_5\right]=-\frac{ie^2}{12\pi^2}\tilde{F}^{0i}
$$
for the axial case and

$$
\left[p^{ai},p^{bj}\right]=\frac{ie^3}{72\pi^2}\varepsilon^{0ijk}A^c_k
d^{abc}
$$
for the chiral one. This way was discussed in ref [27]. Although these
commutation relations are inconsistent by themselves, they do restore the
Wess-Zumino consistency conditions for the generating operators
($J_1=0$, $J_2=0$) for the abelian axial invariant theory.
The explicit value of the anomaly for this case has the form

$$
\left[\Omega,\Omega\right]=0\ ,\ i\left[ H,\Omega\right]=
-\frac{e^2}{24\pi^2}\int^{}_{}d{\bf x}\,{\cal C}_5\tilde{F}^{\alpha\beta}
F_{\alpha\beta}
$$
As to the nonabelian chiral case this way may give only $J_1=0$, while the
second Jacoby identity (5.2) still remains broken. Another way is to
introduce an arbitrary constant multiplier to the connection in the
regularizing function in the current. The hope is to find such a meaning
for the multiplier which may provide both $J_1$ and $J_2$ to be vanished.
This method was discussed in ref [13] with respect to the axial current
anomaly. We can show that it may provide the consistency condition for the
BRST--anomaly in the axial case but the nonabelian chiral case remains
inconsistent again. Thus there are no known means to transform the operator
covariant BRST--anomaly into the consistent one in the nonabelian chiral
theory.

As far as the covariance and the consistency may contradict each other for
an anomalous commutators one may have to choose one of the description
methods: either inconsistent or noncovariant. Our point of view is in the
following: the well defined physical theory should have its anomalies to be
mutually cancelled in the gauge algebra generating equations. Then there
will not appear a problem of this choice, while the anomalous model might
be treated as a part of more wide theory with nilpotent and conserved
BRST-charge. We believe the mutual cancelation of the anomalous
contributions in the generating equations has to be described in the
covariant form and the paper gives a scheme of such a description.

\vspace{10mm}

The authors wish to thank I.A.Batalin for helpful discussions. We are
grateful also to the referee of the paper for a constructive criticism
and very useful suggestions for improving the manuscript.
\vspace{10mm}

The work is partially supported by the ISF long term research grant
No M2I000
and by the European Community grant INTAS-93-2058. One of us (SLL) is also
thankful to the Royal Society for the support under Kapitza Fellowship
Programme.

\newpage

\setcounter{equation}{0}
\setcounter{section}{1}
\renewcommand{\thesection}{\Alph{section}}
\section*{Appendix A}
In appendix A we express the Schwinger terms of the quantum commutators
involved into the constraint algebra, involution relations of the
constraints with the Hamiltonian and in the Heisenberg equations for
the currents in terms of the vacuum expectation values. The vacuum
expectation values are obtained in appendix B.

The basic vacuum expectation value of the bilinear fermion operator has the
following structure:
\begin{eqnarray}
& & V_{\stackrel{{\scriptstyle{l}}}{{\scriptstyle{r}}}}^\mu
({\bf x},t;\mbox{{\boldmath$\varepsilon$}}\vert{\hat{N}}^q)= \\
& &{}=\langle 0\vert\bar{\psi}({\bf x_+},t)\left({\hat{N}}^q
R_{\stackrel{{\scriptstyle{l}}}{{\scriptstyle{r}}}}^{\mu a}({\bf x},t;
\mbox{{\boldmath$\varepsilon$}})\right)\psi({\bf x_-},t)\vert 0
\rangle=\nonumber\\
& &=\bar{\psi}({\bf x_+},t)\left({\hat{N}}^q
R_{\stackrel{{\scriptstyle{l}}}{{\scriptstyle{r}}}} ^{\mu a}
({\bf x},t;\mbox{{\boldmath$\varepsilon$}})\right)\psi
({\bf x_-},t)\nonumber\\
& &\hspace{1mm}\mbox{\hspace{5mm}\rule{0.4pt}{2mm}%
\rule{11.1em}{0.4pt}\rule{0.4pt}{2mm}}\nonumber
\end{eqnarray}
$R_{\stackrel{{\scriptstyle{l}}}{{\scriptstyle{r}}}}^{\mu a}
({\bf x},t;\mbox{{\boldmath$\varepsilon$}})$
is defined by (2.9), ${\hat{N}}^q$ are the operators acting on
$R_{\stackrel{{\scriptstyle{l}}}{{\scriptstyle{r}}}} ^{\mu a}
({\bf x},t;\mbox{{\boldmath$\varepsilon$}})$
only:
\begin{eqnarray}
& &{\hat{N}}^{1b}R_{\stackrel{{\scriptstyle{l}}}{{\scriptstyle{r}}}}^{\mu a}
=  {\left[R_{\stackrel{{\scriptstyle{l}}}{{\scriptstyle{r}}}}^{\mu a},
t^b\right]}_+;\nonumber\\
& &{\hat{N}}^{2b}R_{\stackrel{{\scriptstyle{l}}}{{\scriptstyle{r}}}}^{\mu a}
  = {\left[R_{\stackrel{{\scriptstyle{l}}}{{\scriptstyle{r}}}}^{\mu a},
t^b\right]}_-;\nonumber\\
& &{\hat{N}}^{3b}R_{\stackrel{{\scriptstyle{l}}}{{\scriptstyle{r}}}}^{\mu a}
=  {\left[R_{\stackrel{{\scriptstyle{l}}}{{\scriptstyle{r}}}}^{\mu a},
t^b\right]}_--f^{abc}
R_{\stackrel{{\scriptstyle{l}}}{{\scriptstyle{r}}}}^{\mu c};\\
& &{\hat{N}}^{4}_kR_{\stackrel{{\scriptstyle{l}}}{{\scriptstyle{r}}}}^{ka}
=  {\partial}_k
R_{\stackrel{{\scriptstyle{l}}}{{\scriptstyle{r}}}}^{ka};\nonumber\\
& &{\hat{N}}^{5kb}({\bf y},t)
R_{\stackrel{{\scriptstyle{l}}}{{\scriptstyle{r}}}}^{\mu a}
({\bf x},t;\mbox{{\boldmath$\varepsilon$}})  =
 \frac{\delta}{\delta A^{kb}({\bf y},t)}
R_{\stackrel{{\scriptstyle{l}}}{{\scriptstyle{r}}}}^{\mu a}
({\bf x},t;\mbox{{\boldmath$\varepsilon$}});\nonumber\\
& &{\hat{N}}^{6b}({\bf y},t)
{R_{\stackrel{{\scriptstyle{l}}}{{\scriptstyle{r}}}}}^{\mu a}
({\bf x},t;\mbox{{\boldmath$\varepsilon$}})  =
 \frac{\delta}{\delta \phi^b({\bf y},t)}
R_{\stackrel{{\scriptstyle{l}}}{{\scriptstyle{r}}}}^{\mu a}
({\bf x},t;\mbox{{\boldmath$\varepsilon$}});\nonumber
\end{eqnarray}
According to (2.13) we have:
\begin{equation}
V_{\stackrel{{\scriptstyle{l}}}{{\scriptstyle{r}}}}^{\mu a}
({\bf x},t;\mbox{{\boldmath$\varepsilon$}}\vert{\hat{N}}^q)=
i\,Sp\,Tr\left({\hat{N}}^q
R_{\stackrel{{\scriptstyle{l}}}{{\scriptstyle{r}}}}^{\mu a}
({\bf x},t;\mbox{{\boldmath$\varepsilon$}})
S^c({{\bf x}}_-,t;{{\bf x}}_+,t)\right),
\end{equation}
the trace corresponding to $t^a$ indices is denoted as $Sp$ and the trace
corresponding to the indices of $\gamma$-matrix as $Tr$.

The needed divergent and finite parts of the vacuum expectation values
$V_{\stackrel{{\scriptstyle{l}}}{{\scriptstyle{r}}}}^{\mu a}
({\bf x},t;\mbox{{\boldmath$\varepsilon$}}\vert{\hat{N}}^q)$ (A.3)
are found in appendix B by the proper time expansion of the Green function
(2.14). It is apparent that all the Schwinger terms in the constraint
algebra, in the involution relations of the constraints with the
Hamiltonian and in the Heisenberg equations for the currents can be
expressed in terms of the vacuum expectation values of the following
commutators (or that is the same in terms of their Schwinger terms)
\begin{eqnarray}
{C_{\stackrel{{\scriptstyle{l}}}{{\scriptstyle{r}}}}}^1(g,f;t)&=&
\lim_{\mbox{{\boldmath$\varepsilon$}}\to0}\langle0\vert
\left[\stackrel{l}{\nabla}_ip^i(g,t),
{j_{\stackrel{{\scriptstyle{l}}}{{\scriptstyle{r}}}}}^{0reg}
(f,t;\mbox{{\boldmath$\varepsilon$}})\right]\vert0\rangle,\nonumber\\
{C_{\stackrel{{\scriptstyle{l}}}{{\scriptstyle{r}}}}}^{20}(g,f;t)&=&
\lim_{\stackrel{{\scriptstyle \mbox{{\boldmath$\varepsilon'$}} \to 0}}%
{\mbox{{\boldmath$\varepsilon$}}\to0}}\langle0\vert
\left[{j_{\stackrel{{\scriptstyle{l}}}{{\scriptstyle{r}}}}}^{0reg}
(g,t;\mbox{{\boldmath$\varepsilon$}}),
{j_{\stackrel{{\scriptstyle{l}}}{{\scriptstyle{r}}}}}^{0reg}
(f,t;\mbox{{\boldmath$\varepsilon'$}})\right]\vert0\rangle,\nonumber\\
{C_{\stackrel{{\scriptstyle{l}}}{{\scriptstyle{r}}}}}^{2i}(g,f;t)&=&
\lim_{\mbox{{\boldmath$\varepsilon$}}\to0}\langle0\vert
\left[{j_{\stackrel{{\scriptstyle{l}}}{{\scriptstyle{r}}}}}^{0reg}
(g,t;\mbox{{\boldmath$\varepsilon$}}),
{j_{\stackrel{{\scriptstyle{l}}}{{\scriptstyle{r}}}}}^i(f,t)\right]
\vert0\rangle,\\
{C_{\stackrel{{\scriptstyle{l}}}{{\scriptstyle{r}}}}}^3(g;t)&=&
\lim_{\mbox{{\boldmath$\varepsilon$}}\to0}\langle0\vert i
\left[H_F(t),{j_{\stackrel{{\scriptstyle{l}}}{{\scriptstyle{r}}}}}^{0reg}
(g,t;\mbox{{\boldmath$\varepsilon$}})\right]\vert0\rangle,\nonumber\\
{C_{\stackrel{{\scriptstyle{l}}}{{\scriptstyle{r}}}}}^4(g;t)&=&
\lim_{\mbox{{\boldmath$\varepsilon$}}\to0}\langle0\vert i
\left[H_M(t),{j_{\stackrel{{\scriptstyle{l}}}{{\scriptstyle{r}}}}}^{0reg}
(g,t;\mbox{{\boldmath$\varepsilon$}})\right]\vert0\rangle,\nonumber\\
{C_{\stackrel{{\scriptstyle{l}}}{{\scriptstyle{r}}}}}^5(g,f;t)&=&
\lim_{\mbox{{\boldmath$\varepsilon$}}\to0}\langle0\vert
\left[{j_{\stackrel{{\scriptstyle{l}}}{{\scriptstyle{r}}}}}^{0reg}
(g,t;\mbox{{\boldmath$\varepsilon$}}),p(f,t)\right]\vert0\rangle.\nonumber
\end{eqnarray}
The vacuum expectation values (A.4) in their turn can be expressed in
terms of
$V_{\stackrel{{\scriptstyle{l}}}{{\scriptstyle{r}}}}^{\mu a}
({\bf x},t;\mbox{{\boldmath$\varepsilon$}}\vert{\hat{N}}^q)$
(A.3) by making use of the equal-time commutation relations (2.2), (2.3),
(2.5) only (the coefficients for
$V_{\stackrel{{\scriptstyle{l}}}{{\scriptstyle{r}}}}^{\mu a}
({\bf x},t;\mbox{{\boldmath$\varepsilon$}}\vert{\hat{N}}^q)$
are considered to the third order in $\mbox{{\boldmath$\varepsilon$}}$
because the fermion propagator includes divergences no more than
$\varepsilon^{-3}$):
\begin{eqnarray*}
 & &{C_{\stackrel{{\scriptstyle{l}}}{{\scriptstyle{r}}}}}^{1}(g,f;t)=
\lim_{\mbox{{\boldmath$\varepsilon$}}\to0}\int^{}_{}d{\bf x}\,d{\bf y}\,
f^a({\bf x},t)\left(\stackrel{l}{\nabla}_jg({\bf y};t)\right)^b
V_{\stackrel{\scriptstyle l}{\scriptstyle r}}^{0a}({\bf x},t;
\mbox{{\boldmath$\varepsilon$}}\vert \hat{N}^{5jb}({\bf y},t));\\
 & &C_{\stackrel{\scriptstyle l}{\scriptstyle r}}^{2\mu}(g,f;t)=
\lim_{\mbox{{\boldmath$\varepsilon$}}\to 0}\int^{}_{}d{\bf y}\,g^a
({\bf y},t)\biggl\{ f^b({\bf y},t)f^{abc}
j_{\stackrel{\scriptstyle l}{\scriptstyle r}}^{\mu c\,reg}
({\bf y},t;\mbox{{\boldmath$\varepsilon$}})+\\
 & &\mbox{}+f^b({\bf y},t)
V_{\stackrel{\scriptstyle l}{\scriptstyle r}}^{\mu a}
({\bf y},t;\mbox{{\boldmath$\varepsilon$}}\vert
 \hat{N}^{3b}({\bf y},t))-\frac{1}{2}\mbox{{\boldmath$\varepsilon$}}^j
\partial_jf^b({\bf y},t)
V_{\stackrel{\scriptstyle l}{\scriptstyle r}}^{\mu a}({\bf y},t;
\mbox{{\boldmath$\varepsilon$}}\vert \hat{N}^{1b})+\\
 & &\mbox{}+\frac{1}{8}\mbox{{\boldmath$\varepsilon$}}^j
\mbox{{\boldmath$\varepsilon$}}^k\partial_j\partial_kf^b({\bf y},t)
V_{\stackrel{{\scriptstyle{l}}}{{\scriptstyle{r}}}}^{\mu a}
({\bf y},t;\mbox{{\boldmath$\varepsilon$}}\vert{\hat{N}}^{2b})-\\
 & &\mbox{}-\frac{1}{48}\mbox{{\boldmath$\varepsilon$}}^i
\mbox{{\boldmath$\varepsilon$}}^j\mbox{{\boldmath$\varepsilon$}}^k
\partial_i\partial_j\partial_kf^b({\bf y},t)
{V_{\stackrel{{\scriptstyle{l}}}{{\scriptstyle{r}}}}}^{\mu a}
({\bf y},t;\mbox{{\boldmath$\varepsilon$}}\vert{\hat{N}}^{1b})\biggr\};
\end{eqnarray*}
(by the evaluation of
${C_{\stackrel{{\scriptstyle{l}}}{{\scriptstyle{r}}}}}^{20}(g,f;t)$ (A.4)
the result does not
depend on in which order the local limit of either of the two regularization
parameters is taken). The following notation are further used:
${Z_l}^i=eA^i$,
${Z_r}^i=K^i$.
\begin{eqnarray*}
{C_{\stackrel{{\scriptstyle{l}}}{{\scriptstyle{r}}}}}^{3}(g,t)&=&
\lim_{\mbox{{\boldmath$\varepsilon$}}\to0}\int{}{}d{\bf y}\,
g^a({\bf y},t)\bigg\{  -
\partial_i{j_{\stackrel{{\scriptstyle{l}}}{{\scriptstyle{r}}}}}^{ia\,reg}
({\bf y},t;\mbox{{\boldmath$\varepsilon$}})+\\
& &{}+{V_{\stackrel{{\scriptstyle{l}}}{{\scriptstyle{r}}}}}^{ia}
({\bf y},t;\mbox{{\boldmath$\varepsilon$}}\vert{\hat{N}}^4_i)\bigg\} +
i{C_{\stackrel{{\scriptstyle{l}}}{{\scriptstyle{r}}}}}^{2j}
(g,Z_{\stackrel{{\scriptstyle{l}}}{{\scriptstyle{r}}}}\mbox{}_j;t)
\end{eqnarray*}
(The third term in this expression is the result of the commutation of the
time-component of the regularized current with $H^{int}_F$, the first and
the second ones --- with the free part of the Hamiltonian $H_F$; thus the
both of these commutators have Schwinger terms, being anomalous in the case
of the left current);
\begin{eqnarray*}
& &{C_{\stackrel{{\scriptstyle{l}}}{{\scriptstyle{r}}}}}^{4}(g;t)=
\lim_{\mbox{{\boldmath$\varepsilon$}}\to0}\int{}{}d{\bf x}\,d{\bf y}\,
g^a({\bf x},t)p^b_j({\bf y};t)
{V_{\stackrel{{\scriptstyle{l}}}{{\scriptstyle{r}}}}}^{0a}
({\bf x},t;\mbox{{\boldmath$\varepsilon$}}
\vert{\hat{N}}^{5jb}({\bf y},t));\\
& &{C_{\stackrel{{\scriptstyle{l}}}{{\scriptstyle{r}}}}}^{5}(g;t)=
i\lim_{\mbox{{\boldmath$\varepsilon$}}\to0}\int{}{}d{\bf x}\,
d{\bf y}\,g^a({\bf x},t)f^b({\bf y};t)
{V_{\stackrel{{\scriptstyle{l}}}{{\scriptstyle{r}}}}}^{0a}
({\bf x},t;\mbox{{\boldmath$\varepsilon$}}\vert{\hat{N}}^{6b}({\bf y},t));
\end{eqnarray*}

When substituting the explicit form of the vacuum expectation value
${V_{\stackrel{{\scriptstyle{l}}}{{\scriptstyle{r}}}}}^{\mu a}
({\bf x},t;\mbox{{\boldmath$\varepsilon$}}\vert{\hat{N}}^q)$
from the appendix B
and taking the local limit with the spatial averaging (2.12) we obtain the
following explicit expressions for the desired commutators (A.4):

\begin{eqnarray}
& &C_l{}^1(g,f;t)=\frac{ie^2}{48\pi^2}\int{}{}d{\bf y}\,
Sp\left\{{\tilde{F}}^{0i}({\bf y},t){\left[
\stackrel{l}{\nabla}_ig({\bf y},t),f({\bf y},t)\right]}_+\right\};\\
& &C_r{}^1(g,f;t)=0;\\
& &{C_{\stackrel{{\scriptstyle{l}}}{{\scriptstyle{r}}}}}^{2\mu}(g,f;t)=
\int{}{}d{\bf y}\,\bigg\{ g^a({\bf y},t)f^b({\bf y},t)f^{abc}\langle 0
\vert{j_{\stackrel{{\scriptstyle{l}}}{{\scriptstyle{r}}}}}^{\mu c}
({\bf y},t)\vert0\rangle+\\
& &\mbox{}+\frac{ie}{48\pi^2}Sp\left(g({\bf y},t){\left[
{\tilde{F_{\stackrel{{\scriptstyle{l}}}{{\scriptstyle{r}}}}}}^{\mu i}
({\bf y},t){\stackrel{\stackrel{\scriptstyle l}{\scriptstyle r}}{\nabla}}_i
f({\bf y},t)\right]}_+\right)-\nonumber\\
& &\mbox{}-\frac{i}{\pi^2\mbox{{\boldmath$\varepsilon^4$}}}
\lim_{\mbox{{\boldmath$\varepsilon$}}\to0}\left(
\mbox{{\boldmath$\varepsilon$}}^j\mbox{{\boldmath$\varepsilon$}}^k
\eta^\mu_k\right)%
Sp\left(g({\bf y},t){\stackrel{\stackrel{\scriptstyle
l}{\scriptstyle r}}{\nabla}}_jf({\bf y},t)\right)-\nonumber\\
& &\mbox{}-\frac{i}{24\pi^2\mbox{{\boldmath$\varepsilon^4$}} }
\lim_{\mbox{{\boldmath$\varepsilon$}}\to0}
\left(\mbox{{\boldmath$\varepsilon$}}^j\mbox{{\boldmath$\varepsilon$}}^k
\mbox{{\boldmath$\varepsilon$}}^m\mbox{{\boldmath$\varepsilon$}}^n%
\eta^\mu_n\right)Sp\left(g({\bf y},t){\stackrel{\stackrel{\scriptstyle
l}{\scriptstyle r}}{\nabla}}_j{\stackrel{\stackrel{\scriptstyle
l}{\scriptstyle r}}{\nabla}}_k{\stackrel{\stackrel{\scriptstyle
l}{\scriptstyle r}}{\nabla}}_m%
f({\bf y},t)\right)\bigg\}.\nonumber
\end{eqnarray}

One may see from (A.4),(A.7) that the commutator of current time-component
$j_{\stackrel{\scriptstyle l}{ \scriptstyle r}}^0$ with the interaction part
of the Hamiltonian $H^{int}_F$ (which can be expressed in terms of the
commutator of two currents) involves the contribution which is diverged as
$\mbox{{\boldmath$\varepsilon^{-2}$}}$ and the finite contribution with the
third derivatives. This part of the commutator is shown to cancel completely
when one considers the commutator of
${j_{\stackrel{{\scriptstyle{l}}}{{\scriptstyle{r}}}}}^0$
with the total Hamiltonian
$H_F$, namely:
\begin{eqnarray}
C_{\stackrel{{\scriptstyle{l}}}{{\scriptstyle{r}}}}^{3}(g;t)&=&
\mbox{}-\langle 0 \vert{\stackrel{l}{\nabla}}_i
{j_{\stackrel{{\scriptstyle{l}}}{{\scriptstyle{r}}}}}^{ic}
({\bf y},t)\vert 0\rangle+\\
& &\mbox{}+\frac{e^2}{48\pi^2}\int^{}_{}d{\bf y}\,Sp\left(g
({\bf y},t){\left[
\tilde{F_{\stackrel{{\scriptstyle{l}}}{{\scriptstyle{r}}}}}^{ij}
({\bf y},t),{F_{\stackrel{{\scriptstyle{l}}}{{\scriptstyle{r}}}}}
\mbox{}_{ij}({\bf y},t)\right]}_+\right);\nonumber\\
C^4_l(g;t)&=&\frac{e^2}{48\pi^2}\int^{}_{}d{\bf y}\,Sp\left(g
({\bf y},t){\left[\tilde{F}_l^{0j} ({\bf y},t) , p_j({\bf y},t)\right]}_+
\right),\\
C^4_r(g;t)&=&0;\nonumber\\
{C_{\stackrel{{\scriptstyle{l}}}{{\scriptstyle{r}}}}}^{5\mu}(g,f;t)&=&0.
\end{eqnarray}
The following notations are used here:
\begin{math}
{\tilde{F}}^{\mu\nu}_l={\tilde{F}}^{\mu\nu};\, {\tilde{F}}^{\mu\nu}_r=
{\mbox{{\boldmath$\varepsilon$}}}^{\mu\nu\alpha\beta}F_{r\alpha\beta};\\
F^a_{r\mu\nu}=-\left\{\partial_\mu K^a_\nu - \partial_\nu K^a_\mu -
if^{abc}K^b_\mu K^c_\nu\right\}.
\end{math}
If $K^a_\mu$ are given by (1.6) than $F^a_{r\mu\nu}=0$.
The commutators given below do not contribute to the final expressions
for the Schwinger terms of the current algebra and the involution relations,
 but they are accounted at the intervening calculations thereof.
\begin{eqnarray}
C^6(g;t)  &=&\langle0\vert i\left[H(t),\stackrel{l}{\nabla}_ip^i(g;t)\right]
\vert0\rangle=\\
          &=&e\int^{}_{}d{\bf y}\,g^a({\bf y},t)\langle0\vert
{\left(\stackrel{l}{\nabla}_ij^i_l({\bf y},t)\right)}^a
\vert0\rangle,\nonumber\\
C^7(g;t)  &=&\langle0\vert i\left[H(t),p(g;t)\right]\vert0\rangle=\\
          &=&-\int^{}_{}d{\bf y}\,g^a({\bf y},t)E^{-1ab}({\bf y},t)
\langle0\vert{\left(\stackrel{r}{\nabla}_ij^i_r({\bf y},t)\right)}^b
\vert0\rangle,\nonumber\\
C^8(g,f;t)&=&\langle0\vert\left[\stackrel{l}{\nabla}_ip^i(g;t),
\stackrel{l}{\nabla}_jp^j(g;t)\right]\vert0\rangle=\\
          &=&e\int^{}_{}d{\bf y}\,f^{abc}g^a({\bf y},t)f^b({\bf y},t)
\langle0\vert{\left(\stackrel{l}{\nabla}_ip^i({\bf y},t)\right)}^c
\vert0\rangle.\nonumber
\end{eqnarray}
Thus in (A.5--A.13) the vacuum expectation values of the quantum commutators
(A.4) are explicitly given. They allow to describe completely the quantum
contributions to the constraint algebra and the involution relations.
\setcounter{equation}{0}
\setcounter{section}{2}
\section*{Appendix B}
In appendix B we derive the finite and divergent parts of the vacuum
expectation values
\begin{math}
V_{\stackrel{{\scriptstyle{l}}}{{\scriptstyle{r}}}}^\mu({\bf x},t;
\mbox{{\boldmath$\varepsilon$}}\vert{\hat{N}}^q)
\end{math}
being used to find the quantum commutators (A.4). The vacuum expectation
values are defined as:
\begin{equation}
{V_{\stackrel{{\scriptstyle{l}}}{{\scriptstyle{r}}}}}^\mu({\bf x},t;
\mbox{{\boldmath$\varepsilon$}}\vert{\hat{N}}^q)=
\end{equation}
\begin{displaymath}
{}=\langle 0\vert\bar{\psi}({{\bf x}}_+,t)\left({\hat{N}}^q
{R_{\stackrel{{\scriptstyle{l}}}{{\scriptstyle{r}}}}}^{\mu a}
({\bf x},t;\mbox{{\boldmath$\varepsilon$}})\right)\psi({{\bf x}}_-,t)
\vert 0\rangle=
\end{displaymath}
\begin{displaymath}
{}=i\,Sp\,Tr\left({\hat{N}}^q
{R_{\stackrel{{\scriptstyle{l}}}{{\scriptstyle{r}}}}}^{\mu a}
({\bf x},t;\mbox{{\boldmath$\varepsilon$}})
S^c({{\bf x}}_-,t;{{\bf x}}_+,t)\right),
\end{displaymath}
where ${R_{\stackrel{{\scriptstyle{l}}}{{\scriptstyle{r}}}}}^{\mu a}
({\bf x},t;\mbox{{\boldmath$\varepsilon$}})$ and
$S^c({{\bf x}}_-,t;{{\bf x}}_+,t)$
are given by (2.9) and (2.13); ${\hat{N}}^q$ are the operators of the form
(A.2) acting on
${R_{\stackrel{{\scriptstyle{l}}}{{\scriptstyle{r}}}}}^{\mu a}
({\bf x},t;\mbox{{\boldmath$\varepsilon$}})$
only. The equations for fermion propagator $S^c(x,y)$ (2.14) and for the
${R_{\stackrel{{\scriptstyle{l}}}{{\scriptstyle{r}}}}}^{\mu a}
({\bf x},t;\mbox{{\boldmath$\varepsilon$}})$ (2.9)
use the connection which includes the additional fields ${\phi}^a$.

The vacuum expectation value is derived in the following order. All
functions in (B.1) are represented as the expansion in terms of
$\varepsilon$ with the coefficients depending on $({\bf x},t)$ only.
Then the divergent and finite parts of (B.1) are extracted.

The expansion of ${R_{\stackrel{{\scriptstyle{l}}}{{\scriptstyle{r}}}}}^{\mu
a}({\bf x},t;\mbox{{\boldmath$\varepsilon$}})$ (2.9) in terms of
$\varepsilon$
is of the form:
\begin{eqnarray}
R_{\stackrel{{\scriptstyle{l}}}{{\scriptstyle{r}}}}^{\mu a}
({\bf x},t;\mbox{{\boldmath$\varepsilon$}})&=&\gamma^\mu
P_{\stackrel{{\scriptstyle{l}}}{{\scriptstyle{r}}}}\bigg\{ t^a+
b^{(1)a}{}_{i}\varepsilon^i+b^{(2)a}{}_{ij}\varepsilon^i\varepsilon^j+\\
& &\mbox{}+b^{(3)a}{}_{ijk}\varepsilon^i\varepsilon^j\varepsilon^k+
\sim{\varepsilon}^4\bigg\} ,\nonumber
\end{eqnarray}
where
\begin{eqnarray*}
& &b^{(1)a}{}_{i}=\frac{i}{2}{\left[\Gamma_i,t^a\right]}_+;\\
& &b^{(2)a}{}_{ij}=-\frac{1}{8}{\left[\Gamma_i\Gamma_j,t^a\right]}_+-
\frac{i}{8}{\left[t^a,\partial_i\Gamma_j\right]}_--\frac{1}{4}\Gamma_i
t^a\Gamma_j;\\
& &b^{(3)a}{}_{ijk}=\frac{i}{48}{\left[\partial_i\partial_j\Gamma_k,
t^a\right]}_+-\frac{i}{48}{\left[\Gamma_i\Gamma_j\Gamma_k,t^a\right]}_+-\\
& &-\frac{1}{32}{\left[\partial_i\Gamma_j\Gamma_k+\Gamma_k\partial_i
\Gamma_j,t^a\right]}_-+\frac{1}{16}\left(\Gamma_it^a\partial_j\Gamma_k-
\partial_i\Gamma_jt^a\Gamma_k\right)-\\
& &-\frac{i}{16}\left(\Gamma_it^a\Gamma_j\Gamma_k+\Gamma_i\Gamma_j
t^a\Gamma_k\right)+\frac{1}{48}t^a\left[\Gamma_i,\partial_j
\Gamma_k\right]_-.
\end{eqnarray*}
The expansion need comprise terms up to $\varepsilon^3$ only because
the fermion propagator diverges as no more than $\varepsilon^{-3}$.

The representation of $S^c({{\bf x}}_-,t;{{\bf x}}_+,t)$ as a power series
in $\varepsilon$ is done in such a way. The fermion propagator
$S^c(y_-,y_+)$ is expressed as a power series in $\varepsilon$ with the
coefficient depending on $y$ only. Then the divergent and finite parts
are separated and the limit $\varepsilon^0\to0$ is taken. The resulting
expression is substituted into (B.1).

The derivation of the finite and divergent parts of the fermion propagator
$S^c(y_-,y_+)$ is done covariantly by making use Schwinger deWitt
technique \cite{dew} which does not have a need to fix a gauge for the
vector field when the fermions are under the consideration.

Let us $S^c(y_-,y_+)$ is represented as
\begin{equation}
S^c(y_-,y_+)=-i\gamma^\nu\nabla_\nu{\cal G}^c(y_-,y_+),
\end{equation}
As is clear from (2.14) ${\cal G}^c(y_-,y_+)$ is subject to the equation:
\begin{equation}
\gamma^\nu\nabla_\nu\gamma^\alpha\nabla_\alpha{\cal G}^c(y_-,y_+)=-
\delta(y_-,y_+).
\end{equation}
The covariant differentiation the two-point function in this paper is
carried out with respect to the left argument only. The solution of (B.4)
is expressed in form:
\begin{eqnarray}
& &{\cal G}^c(y_-,y_+)=\frac{1}{16\pi^2}\int_{0}^{\infty}\frac{ds}{s^2}
(is)^n\exp\left[\frac{i\sigma}{2s}\right] a_n(y_-,y_+),\\
& &\sigma=\frac{1}{2}(y_--y_+)^2=\frac{1}{2}\varepsilon^2; \ \sigma_\alpha
=(y_--y_+)_\alpha.\nonumber
\end{eqnarray}
The coefficients $a_n(y_-,y_+)$ obey the following recurrent relations:
\begin{eqnarray}
& &\gamma^\alpha\gamma^\beta\sigma_\alpha\nabla_\beta a_0(y_-,y_+)+
\gamma^\alpha\sigma_\beta\nabla_\alpha\gamma^\beta a_0(y_-,y_+)=0;\\
& &\frac{1}{2}\gamma^\alpha\gamma^\beta\sigma_\alpha\nabla_\beta
a_n(y_-,y_+)+\frac{1}{2}\gamma^\alpha\sigma_\beta\nabla_\alpha\gamma^\beta
a_n(y_-,y_+)+\nonumber\\
& &{}+na_n(y_-,y_+)=\gamma^\alpha\nabla_\alpha\gamma^\beta\nabla_\beta
a_{n-1}(y_-,y_+);\nonumber\\
& &\lim_{\varepsilon\to 0}a_0(y_-,y_+)=1.\nonumber
\end{eqnarray}
It should be mentioned that the covariant derivatives in (B.6) do not
commute with $\gamma$--matrix. According to (B.3) and (B.5) the divergent
and finite parts of $S^c(y_-,y_+)$ can be represented as:
\begin{eqnarray}
& &S^c(y_-,y_+)=\frac{1}{16\pi^2}\gamma^\nu\bigg\{ 8
\frac{{\varepsilon}_\nu}{\varepsilon^4}a_0(y_-,y_+)+
\frac{4}{\varepsilon^2}\nabla_\nu a_0(y_-,y_+)+\\
& &+2\frac{\varepsilon_\nu}{\varepsilon^2}a_1(y_-,y_+)+
\nabla_\nu a_1(y_-,y_+)\int^\infty_0\frac{ds}{s}
\exp\left(\frac{i\sigma}{2s}\right)\bigg\} +\sim\varepsilon.\nonumber
\end{eqnarray}
First, all the two-points functions in (B.7) are represented as a power
series in $\varepsilon$ with the coefficients depending on $\Gamma(y)$ only
(the differentiation the recurrent relations (B.6) and subsequent limit
$\varepsilon\to 0$ give all the desired derivatives of
$a_n(y_-,y_+)$ up to second order). Then, when setting
$\varepsilon^0\to 0$, the resulting expression is substituted to (B.1).

After taking the trace $Tr$ of all the products of $\gamma$-matrix in
${V_{\stackrel{{\scriptstyle{l}}}{{\scriptstyle{r}}}}}^\mu
({\bf x},t;\mbox{{\boldmath$\varepsilon$}}\vert{\hat{N}}^q)$
(B.1) we obtain (with the accuracy to those order in $\varepsilon$ which
is used to define the quantum commutators (A.4)):
\begin{eqnarray}
& &V_{\stackrel{{\scriptstyle{l}}}{{\scriptstyle{r}}}}^{\mu a}
({\bf x},t;\mbox{{\boldmath$\varepsilon$}}\vert {\hat{N}}^{1b})=
\frac{i\varepsilon^i\eta^\mu{}_i}{\pi^2\mbox{{\boldmath$\varepsilon$}}^4}
Sp{\left[ t^a,t^b \right]}_++\\
& &+\frac{\varepsilon_i\varepsilon_j\varepsilon_k\eta^{\mu i}}{4\pi^2
{\mbox{{\boldmath$\varepsilon$}}}^4}Sp\bigg\{ -it^a{\left[
{Z_{\stackrel{{\scriptstyle{l}}}{{\scriptstyle{r}}}}}^j,
{\left[{Z_{\stackrel{{\scriptstyle{l}}}{{\scriptstyle{r}}}}}^k,t^a\right]}_-
\right]}_-+\nonumber\\
& &+t^a{\left[\partial^j
{Z_{\stackrel{{\scriptstyle{l}}}{{\scriptstyle{r}}}}}^k,t^a\right]}_-
\bigg\} -\frac{ie\varepsilon_i}{8\pi^2{\mbox{{\boldmath$\varepsilon$}}}^2}
Sp\left\{{\left[t^a,t^b\right]}_+
{\tilde{F_{\stackrel{{\scriptstyle{l}}}{{\scriptstyle{r}}}}}}^{\mu i}
\right\};\nonumber\\
& &{V_{\stackrel{{\scriptstyle{l}}}{{\scriptstyle{r}}}}}^{\mu a}
({\bf x},t;\mbox{{\boldmath$\varepsilon$}}\vert{\hat{N}}^{2b})=
\frac{i\varepsilon_i\varepsilon_j\eta^{\mu i}}{\pi^2
{\mbox{{\boldmath$\varepsilon$}}}^2}Sp\left\{t^a{\left[t^b,
{Z_{\stackrel{{\scriptstyle{l}}}{{\scriptstyle{r}}}}}^j\right]}_-\right\}-\\
& &-\frac{ie\varepsilon_i}{8\pi^2{\mbox{{\boldmath$\varepsilon$}}}^2}
Sp\left\{{\left[t^a,t^b\right]}_-
{\tilde{F_{\stackrel{{\scriptstyle{l}}}{{\scriptstyle{r}}}}}}^{\mu i}
\right\};\nonumber\\
& &V_{\stackrel{{\scriptstyle{l}}}{{\scriptstyle{r}}}}^{\mu a}
({\bf x},t;\mbox{{\boldmath$\varepsilon$}}\vert {\hat{N}}^{3b})=
-\frac{i\varepsilon_i\varepsilon_j\eta^{\mu i}}{\pi^2
\mbox{{\boldmath$\varepsilon$}}^2}Sp\left\{ t^a{\left[
Z_{\stackrel{{\scriptstyle{l}}}{{\scriptstyle{r}}}}^j,t^b \right]}_-\right\}
 +\\
& &+\frac{e\varepsilon_i\varepsilon_j}{16\pi^2
\mbox{{\boldmath$\varepsilon$}}^2}Sp\left\{ t^a{\left[
\tilde{F_{\stackrel{{\scriptstyle{l}}}{{\scriptstyle{r}}}}}^{\mu j},
{\left[ Z_{\stackrel{{\scriptstyle{l}}}{{\scriptstyle{r}}}}^i,
t^b \right]}_- \right]}_+ \right\} +\nonumber\\
& &+\frac{\varepsilon_i\varepsilon_j\varepsilon_k\varepsilon_l
\eta^{\mu l}}{{\mbox{{\boldmath$\varepsilon$}}}^4}Sp\bigg\{ -
\frac{1}{24\pi^2}t^a{\left[t^b,
{Z_{\stackrel{{\scriptstyle{l}}}{{\scriptstyle{r}}}}}^i
{Z_{\stackrel{{\scriptstyle{l}}}{{\scriptstyle{r}}}}}^j
{Z_{\stackrel{{\scriptstyle{l}}}{{\scriptstyle{r}}}}}^k
\right]}_-+\nonumber\\
& &+\frac{1}{8\pi^2}t^a
{Z_{\stackrel{{\scriptstyle{l}}}{{\scriptstyle{r}}}}}^i{\left[t^b,
{Z_{\stackrel{{\scriptstyle{l}}}{{\scriptstyle{r}}}}}^j\right]}_-
{Z_{\stackrel{{\scriptstyle{l}}}{{\scriptstyle{r}}}}}^k+
\frac{1}{24\pi^2}t^a{\left[t^b,\partial^i\partial^j
{Z_{\stackrel{{\scriptstyle{l}}}{{\scriptstyle{r}}}}}^k
\right]}_--\nonumber\\
& &-\frac{i}{16\pi^2}t^a{\left[
{Z_{\stackrel{{\scriptstyle{l}}}{{\scriptstyle{r}}}}}^i,
{\left[t^b,\partial^j
{Z_{\stackrel{{\scriptstyle{l}}}{{\scriptstyle{r}}}}}^k\right]}_-
\right]}_--\nonumber\\
& &-\frac{i}{16\pi^2}t^a{\left[\partial^i
{Z_{\stackrel{{\scriptstyle{l}}}{{\scriptstyle{r}}}}}^j,
{\left[t^b,{Z_{\stackrel{{\scriptstyle{l}}}{{\scriptstyle{r}}}}}^k\right]}_-
\right]}_--\nonumber\\
& &-\frac{i}{48\pi^2}t^a{\left[t^b,{\left[
{Z_{\stackrel{{\scriptstyle{l}}}{{\scriptstyle{r}}}}}^i,
\partial^j{Z_{\stackrel{{\scriptstyle{l}}}{{\scriptstyle{r}}}}}^k\right]}_-
\right]}_-\bigg\} ;\nonumber\\
& &{V_{\stackrel{{\scriptstyle{l}}}{{\scriptstyle{r}}}}}^{ia}
({\bf x},t;\mbox{{\boldmath$\varepsilon$}}\vert{\hat{N}}^4_i)=
-\frac{\varepsilon^i\varepsilon^j}{\pi^2
{\mbox{{\boldmath$\varepsilon$}}}^4}Sp{\left[t^a\partial^i
{Z_{\stackrel{{\scriptstyle{l}}}{{\scriptstyle{r}}}}}^j\right]}_++\\
& &+\frac{e\varepsilon_i\varepsilon_j}{16\pi^2
{\mbox{{\boldmath$\varepsilon$}}}^2}Sp\left\{t^a
{\left[{\tilde{F_{\stackrel{{\scriptstyle{l}}}{{\scriptstyle{r}}}}}}^{ki},
\partial_k{Z_{\stackrel{{\scriptstyle{l}}}{{\scriptstyle{r}}}}}^j\right]}_+
\right\}+\nonumber\\
& &+\frac{\varepsilon_i\varepsilon_j\varepsilon_k
\varepsilon_l}{{\mbox{{\boldmath$\varepsilon$}}}^4}Sp\bigg\{ -
\frac{1}{24\pi^2}t^a\partial^i\partial^j\partial^k
{Z_{\stackrel{{\scriptstyle{l}}}{{\scriptstyle{r}}}}}^l+\nonumber\\
& &+\frac{1}{24\pi^2}t^a{\left[
{Z_{\stackrel{{\scriptstyle{l}}}{{\scriptstyle{r}}}}}^i,{\left[
{Z_{\stackrel{{\scriptstyle{l}}}{{\scriptstyle{r}}}}}^j,
\partial^k{Z_{\stackrel{{\scriptstyle{l}}}{{\scriptstyle{r}}}}}^l\right]}_-
\right]}_-+\nonumber\\
& &+\frac{i}{12\pi^2}t^a{\left[
{Z_{\stackrel{{\scriptstyle{l}}}{{\scriptstyle{r}}}}}^i,\partial^j
\partial^k{Z_{\stackrel{{\scriptstyle{l}}}{{\scriptstyle{r}}}}}^l\right]}_-
\bigg\} .\nonumber\\
& &V_{\scriptstyle{l}}^{0a}({\bf y},t;\mbox{{\boldmath$\varepsilon$}}
\vert{\hat{N}}^{5jb}({\bf x},t))=\\
& &=\frac{e^2\varepsilon_i\varepsilon_j}{16\pi^2{\mbox{{\boldmath$
\varepsilon$}}}^2}Sp\left\{{\left[t^a,t^b\right]}_+
\tilde{F_{\scriptstyle{l}}}^{0i}\right\}\delta({\bf x}-{\bf y});\nonumber\\
& &V^{0a}_r({\bf y},t;\mbox{{\boldmath$\varepsilon$}}
\vert{\hat{N}}^{5jb}({\bf x},t))=
{V_{\stackrel{{\scriptstyle{l}}}{{\scriptstyle{r}}}}}^{0a}
({\bf x},t;\mbox{{\boldmath$\varepsilon$}}\vert{\hat{N}}^{6b}({\bf x},t))=0.
\end{eqnarray}
When the theory with the covariant derivative (1.2) is studied one should
put $\phi=0$ in (B.8--B.13). In the theory with the axial anomaly all
the vacuum expectation values of the regularized currents
\begin{displaymath}
V^\nu_{(5)}({\bf x},t;\mbox{{\boldmath$\varepsilon$}})=
\langle0\vert j^\nu_{(5)}({\bf x},t;\mbox{{\boldmath$\varepsilon$}})
\vert0\rangle
\end{displaymath}
are derived by the above method (all the comments upon the vacuum
and normal ordering remain valid for this case). The following results are
obtained for the vacuum expectation values to the order which contributes
the desired quantum commutators:
\begin{equation}
V^\nu({\bf x},t;\mbox{{\boldmath$\varepsilon$}})=-
\frac{2\delta^\nu_k\varepsilon^k}{i\pi^2
{\mbox{{\boldmath$\varepsilon$}}}^4}\ ;\ V^\nu_5
({\bf x},t;\mbox{{\boldmath$\varepsilon$}})=
\frac{e\varepsilon_i}{4i\pi^2
{\mbox{{\boldmath$\varepsilon$}}}^2}
{\tilde{F_{\stackrel{{\scriptstyle{l}}}{{\scriptstyle{r}}}}}}^{\nu
 i}({\bf x},t).
\end{equation}
The vacuum expectation values obtained in appendix B define the commutators
considered in appendix A.

\end{document}